\documentstyle[12pt]{article}
\begin{document}

\title{On Wilson Criterion}

\author{Yury M.~Zinoviev\thanks{Supported by the
Russian Foundation of Fundamental Researches under Grant 93-011-147.}\\
Steklov Mathematical Institute, Gubkin St. 8,\\
Moscow 117966, GSP-1, Russia\\
e-mail: zinoviev@genesis.mi.ras.ru }

\date{}
\maketitle

\noindent 
$U(1)$ gauge theory with  the Villain action on a cubic lattice
approximation of three- and four-dimensional torus is considered. 
The naturally chosen correlation functions converge to the correlation 
functions of the ${\bf R}$-gauge electrodynamics on three- and 
four-dimensional torus as the lattice spacing approaches zero
only for the special scaling. This special scaling depends on a choice
of a correlation function system. Another scalings give the degenerate
continuum limits. The Wilson criterion for the confinement is
ambiguous. The asymptotics of the smeared Wilson loop integral for the
large loop perimeters is defined by the density of the loop smearing
over a torus which is transversal to the loop plane. 
When the initial torus radius tends to infinity the correlation
functions converge to the correlation functions of the ${\bf R}$-gauge
Euclidean electrodynamics.

\vskip 1cm

\noindent K. Wilson \cite{1} related the confinement problem to the
study of the correlation functions for the lattice pure gauge theories.
The bulk of the paper \cite{1} is devoted to $U(1)$ gauge theory with 
the periodic boundary conditions. The Wilson criterion \cite{1} for the
confinement is fulfilled if for the large closed loops $\Gamma $ the
correlation function $\langle \exp \{ i \theta (\Gamma )\} \rangle $
looks like $\exp \{ - a \Sigma (\Gamma )\} $ where $\Sigma (\Gamma )$
is the minimal area of a surface with boundary $\Gamma $.
        
This paper is concerned with the case where the gauge group is 
$U(1) = {\bf R}/{(2\pi {\bf Z})}$. Let $h(\theta )$ be a real twice
continuously differentiable even periodic function with period $2\pi $.
The main examples of interest are the Wilson \cite{1} energy function
$h(\theta ) = 1 - \cos \theta $ and the Villain \cite{2} energy  function
\begin{equation}
\label{1}
\exp [-\beta h_{\beta } (\theta )] = c_{\beta } \sum_{n = -\infty }^{\infty }
\exp [-\beta {(\theta - 2\pi n)^{2} }/2]
\end{equation}
where $\beta > 0$ and $c_{\beta } $ is a constant chosen such that the
right-hand side is one for $\theta = 0$.

Let $e_{i} $, $i = 1,...,d$ be the standard unit vectors in ${\bf R}^{d} $,
and $p$ be a non-negative integer less than $d$. The $p$-cells based
at ${\bf m} \in {\bf Z}^{d} $ are the formal symbols:
$({\bf m}; e_{i_{1}} ,...,e_{i_{p}} )$ where the unit vectors differ
from each other.

Let $G$ be one of three abelian groups: ${\bf Z}$, ${\bf R}$, or
$U(1) = {\bf R}/{(2\pi {\bf Z})}$. A $p$-cochain with the coefficients
in $G$ is a $G$-valued function on $p$-cells
$f({\bf m};e_{i_{1}} ,...,e_{i_{p} } ) \equiv f_{i_{1} \cdots i_{p} }
({\bf m})$ which is antisymmetric under the permutations of
the indices $i_{1} ,...,i_{p} $. Let $\Lambda =
\{ {\bf m} \in {\bf Z}^{d} : N_{1} \leq m_{i} \leq N_{2} , i = 1,...,d \} $
be a cube in ${\bf Z}^{d} $ for some integers $N_{1} $ and
$N_{2}$. The conditions

$$
f_{i_{1} \cdots i_{p} } (m_{1} ,...,m_{j} + N,...,m_{d} ) =
f_{i_{1} \cdots i_{p} } ({\bf m}),
$$
where $N = N_{2} - N_{1} + 1$, for every $j = 1,...,d$ and ${\bf m} \in
{\bf Z}^{d} $ correspond to the choice of periodic boundary conditions.
For the periodic boundary conditions we define the boundary operator
\begin{equation}
\label{3}
(\partial f)_{i_{1} \cdots i_{p - 1} } ({\bf m}) = \sum_{\epsilon = 0,1}
\sum_{i_{0} = 1}^{d} (- 1)^{\epsilon + 1}
f_{i_{0} i_{1} \cdots i_{p - 1} } ({\bf m}  - \epsilon e_{i_{0} } )
\end{equation}
and the coboundary operator
\begin{equation}
\label{4}
(\partial^{\ast } f)_{i_{1} \cdots i_{p + 1} } ({\bf m}) =
\sum_{\epsilon = 0,1} \sum_{k = 1}^{p + 1} (- 1)^{\epsilon + k}
f_{i_{1} \cdots \widehat{i_{k} } \cdots i_{p + 1} }
({\bf m} + \epsilon e_{i_{k} } ).
\end{equation}

For the $p$-cochains with coefficients in ${\bf Z}$ or ${\bf R}$ the
inner product is defined by
\begin{equation}
\label{5}
(f,g) = \sum_{i_{1} < \cdots  < i_{p} } \sum_{{\bf m} \in \Lambda  }
f_{i_{1} \cdots i_{p} } ({\bf m}) g_{i_{1} \cdots i_{p} } ({\bf m}) .
\end{equation}

The energy function $h$ and $1$-cochain $\theta $  on $\Lambda $
with coefficients $U(1)$
provide the $2$-cochain on $\Lambda $ with real coefficients.
This $2$-cochain is defined for any indices $i_{1} < i_{2} $ 
by the following relation
$(h(\partial^{\ast} \theta ))_{i_{1} i_{2} } ({\bf m})
\equiv h((\partial^{\ast } \theta )_{i_{1} i_{2} } ({\bf m}))$.
By $1$ we denote $2$-cochain $(1)_{i_{1} i_{2} } ({\bf m}) = 1$ 
for any indices $i_{1} < i_{2} $.

The finite volume Gibbs state in a cube $\Lambda \subset {\bf Z}^{d} $,
at inverse temperature $\beta $ and with energy function $h_{\beta }$ 
is given by
\begin{equation}
\label{6}
\langle F \rangle_{\Lambda ,\beta } =  Z^{- 1} \biggl[
\prod_{{\bf m} \in \Lambda ; i = 1,...,d} \int_{- \pi }^{\pi }
d\theta_{i} ({\bf m}) \biggr] F(\theta )
\exp \bigl[ - \beta (h_{\beta } (\partial^{\ast } \theta ),1) \bigr].
\end{equation}
Here $\theta $ is a $1$-cochain on $\Lambda $ with coefficients  in $U(1)$
and $\theta $ satisfies the periodic boundary conditions.
The measure $d\theta_{i} ({\bf m})$ is Lebesgue measure on
$[- \pi ,\pi ]$. $Z$ is the normalization  constant and $F$ is a
function of the bond variables $\theta_{i} ({\bf m})$.

This paper is concerned with the case of the Villain energy function and
the periodic boundary conditions. We  study the correlation functions:
$\langle  \exp [i(j,\theta )] \rangle_{\Lambda ,\beta } $ where $j$ is a
$1$-cochain on $\Lambda $ with the integer coefficients. The inner
product $(j,\theta )$ is not defined for a $1$-cochain $\theta $ on
$\Lambda $ with coefficients in $U(1) = {\bf R}/{(2\pi {\bf Z})}$,
but $\exp [i(j,\theta )]$ is well defined. It is easy to show that
$\langle  \exp [i(j,\theta )] \rangle_{\Lambda ,\beta } = 0$ if
$j \neq \partial \phi $ for  some $2$-cochain  $\phi $ on $\Lambda $
with  the integer coefficients  (see, for example, \cite{3} ). In view of
the  periodic boundary conditions we can identify the opposite vertices
of the cube $[N_{1},N_{2} + 1]^{\times d} $ and obtain a lattice
approximation ${\bf T}_{N}^{d} $ of the torus ${\bf T}^{d} $ of radius $R$,
where $N = N_{2} - N_{1} + 1$. The $p$-cochains with the  coefficiens in 
the abelian group $G = {\bf Z}, {\bf R} ,U(1) = {\bf R}/{(2\pi {\bf Z})}$ 
satisfying the periodic boundary conditions form the abelian group
$C^{p} ({\bf T}_{N}^{d} ,G)$.

Let $f_{i_{1} \cdots i_{p} } ({\bf x})$ be the coefficients of a real
smooth differential $p$-form on the torus ${\bf T}^{d} $. The inner
product of the differential forms is similar to the inner product
(\ref {5}). We define the differential operator on the differential
forms
\begin{equation}
\label{7}
(df)_{i_{1} \cdots i_{p + 1} } ({\bf x}) = \sum_{k = 1}^{p + 1}
(- 1)^{k + 1} \frac{\partial }{\partial x_{i_{k}}}
f_{i_{1} \cdots \widehat{i_{k} } \cdots i_{p + 1} }({\bf x})
\end{equation}
and its adjoint operator
\begin{equation}
\label{8}
(d^{\ast } f)_{i_{1} \cdots i_{p - 1} } ({\bf x}) = -
\sum_{i_{0} = 1}^{d} \frac{\partial }{\partial x_{i_{0}}}
f_{i_{0} i_{1} \cdots i_{p - 1} } ({\bf x}).
\end{equation}
We define the integer valued $p$-cochain on ${\bf T}_{N}^{d} $
\begin{equation}
\label{9}
(f_{N,b} )_{i_{1} \cdots i_{p} } ({\bf m}) =
[N^{b} f_{i_{1} \cdots i_{p} } (2\pi R N^{-1} {\bf m})],
\end{equation}
where $N = N_{2} - N_{1} + 1$, $b$ is a strictly positive integer, and
$[r]$ is the integer part of the real number $r$. 

The definitions (\ref{3}), (\ref{8}), and (\ref{9}) imply that
\begin{equation}
\label{10}
\lim_{N \rightarrow \infty } (2\pi R)^{- 1} N^{- b + 1}
(\partial f_{N,b} )_{i_{1} \cdots i_{p - 1} } ({\bf m}(N)) =
(d^{\ast } f)_{i_{1} \cdots i_{p - 1} } ({\bf x})
\end{equation}
where $2\pi R N^{-1} {\bf m}(N)$ tends to ${\bf x}$ as
$N \rightarrow \infty $.

\noindent {\bf Proposition 1.} {\it Let a correlation function}
$
\langle \exp [i(\partial \phi_{N,b} ,\theta )]\rangle_{{\bf T}_{N}^{d},\beta}
$ {\it be given by the equality} (\ref{6}). 
{\it Then, for any real smooth differential} 2 - {\it form}
$\phi $ {\it on the torus} ${\bf T}^{d}$, {\it such that}
$d^{\ast }\phi \not\equiv 0$, {\it and, for any numbers} $\beta_{0} > 0$,
$\gamma < d + 2b$, {\it we have}
\begin{equation}
\label{11}
\lim_{N \rightarrow \infty } 
\langle \exp [i(\partial \phi_{N,b} ,\theta )]
\rangle_{{\bf T}_{N}^{d},\beta_{0} N^{\gamma }} = 0.
\end{equation}
{\it Proof}. Let $z_{1} ,...,z_{g} $ form a basis of the group of 
$2$-cycles $Z_{2} ({\bf T}_{N}^{d} ,{\bf Z})$ which is the kernel of the 
homomorphism $\partial : C^{2} ({\bf T}_{N}^{d} ,{\bf Z}) \rightarrow
C^{1} ({\bf T}_{N}^{d} ,{\bf Z})$. The symmetric $g \times g$ matrix
$\Omega_{ij} =  (z_{i} ,z_{j} )$ is positively definite and invertible.
Let us introduce the dual basis
${\bar{z} }_{i} = \sum_{j = 1}^{g} \Omega_{ij}^{- 1} z_{j} $.
Due to \cite{4} the following equality holds
\begin{equation}
\label{12}
\langle \exp [i(\partial \phi_{N,b} ,\theta )]\rangle_{{\bf T}_{N}^{d},\beta}
= W_{{\bf T}_{N}^{d} ,\beta } (\partial \phi_{N,b} )
\frac{\Theta ((\phi_{N,b} ,\bar{{\bf z}} )| 2\pi i\beta \Omega^{- 1} )}{
\Theta ({\bf 0}| 2\pi i\beta \Omega^{- 1} )},
\end{equation}
where
\begin{equation}
\label{13}
W_{{\bf T}_{N}^{d} ,\beta } (\partial \phi_{N,b} ) = 
\exp \bigl[ - 1/{2\beta }[(\phi_{N,b} ,\phi_{N,b} ) - 
\sum_{i = 1}^{g} (\phi_{N,b} ,z_{i})(\phi_{N,b} ,{\bar{z} }_{i})]\bigr]
\end{equation}
is the correlation function calculated for ${\bf R}$-gauge electrodynamics
on the lattice ${\bf T}_{N}^{d} $ in the paper \cite{5} and
the Riemann $\theta $-function
\begin{equation}
\label{14}
\Theta ({\bf y}|\omega ) = \sum_{{\bf m} \in {\bf Z}^{g} }
\exp \bigl[ i\pi \sum_{j,k = 1}^{g} m_{j} \omega_{jk} m_{k} +
2\pi i \sum_{j = 1}^{g} m_{j} y_{j} \bigr]
\end{equation}
depends on the vector ${\bf y} \in {\bf C}^{g} $ and on the symmetric
$g \times g$ matrix $\omega $ with the positively definite imaginary part.
In our case, $\omega = 2\pi i\beta \Omega^{- 1} $.

Relation (\ref{12}) and definition (\ref{14}) imply the
following inequality
\begin{equation}
\label{15}
|
\langle \exp [i(\partial \phi_{N,b} ,\theta )]\rangle_{{\bf T}_{N}^{d},\beta}
| \leq W_{{\bf T}_{N}^{d} ,\beta } (\partial \phi_{N,b} ).
\end{equation}
In \cite{4}, the following limit is calculated:
\begin{eqnarray}
\label{16}
& &\lim_{N \rightarrow \infty } 
N^{- d - 2b}\bigl[ (\phi_{N,b} ,\phi_{N,b} ) -
\sum_{i = 1}^{g} (\phi_{N,b} ,z_{i})(\phi_{N,b} ,{\bar{z} }_{i})\bigr] =
\nonumber \\
& &(2\pi R)^{- 2d} \sum_{\mu = 1}^{d}
\sum_{{\bf l} \in {\bf Z}^{d} ; l_{1}^{2} + \cdots + l_{d}^{2} \neq 0}  
R^{2}(l_{1}^{2} + \cdots + l_{d}^{2} )^{- 1}
|(d^{\ast } \phi )_{\mu }^{\sim }({\bf l})|^{2},
\end{eqnarray}
where
\begin{equation}
\label{17}
(f)_{\mu }^{\sim }({\bf l}) =
\int_{0}^{2\pi R} d\theta_{1} \cdots \int_{0}^{2\pi R} d\theta_{d} 
\exp \bigl[ iR^{- 1} \sum_{k = 1}^{d} l_{k} \theta_{k}  \bigr] 
(f)_{\mu }(\mbox{\boldmath{$\theta$}}).
\end{equation}
Since $d^{\ast }\phi \not\equiv 0$, the limit (\ref{16}) is a 
positive number. Now, the relation (\ref{11}) follows from  
inequality (\ref{15}) and relation (\ref{16}).

\noindent {\bf Proposition 2.} {\it Let a correlation function}
$
\langle \exp [i(\partial \phi_{N,b} ,\theta )]\rangle_{{\bf T}_{N}^{d},\beta}
$ {\it be given by the equality} (\ref{6}). 
{\it Then, for any real smooth differential} 2 - {\it form}
$\phi $ {\it on the torus} ${\bf T}^{d}$, $d = 3,4$, 
{\it and for any numbers} $\beta_{0} > 0$, $\gamma > d + 2b$, {\it we have}
\begin{equation}
\label{18}
\lim_{N \rightarrow \infty } 
\langle \exp [i(\partial \phi_{N,b} ,\theta )]
\rangle_{{\bf T}_{N}^{d},\beta_{0} N^{\gamma }} = 1.
\end{equation}
{\it Proof}. Proposition 2.4 and Proposition 3.4 in \cite{4} imply
that for any sequence $\phi_{N} \in C^{2} ({\bf T}_{N}^{d} ,{\bf Z})$ 
and for any numbers $\beta_{0} > 0$, $\alpha > d$, $d = 3,4$, we have

\begin{equation}
\label{19}
\lim_{N \rightarrow \infty }
\Theta ((\phi_{N} ,\bar{{\bf z}} )| 2\pi i\beta_{0} N^{\alpha }\Omega^{- 1} )
 = 1.
\end{equation}
Now equality (\ref{18}) follows from equalities (\ref{12}),
(\ref{13}), (\ref{16}), and (\ref{19}).

We note that the right-hand side of equality (\ref{16}) coincides
with $(2\pi R)^{- d}(d^{\ast }\phi ,G(d^{\ast }\phi ))$, where
operator $G$ is the Green's operator for the Laplace-Beltrami
operator on the differential $1$-forms on the torus ${\bf T}^{d} $.
In \cite{4}, the following proposition is proved.

\noindent {\bf Proposition 3.} {\it Let a correlation function}
$
\langle \exp [i(\partial \phi_{N,b} ,\theta )]\rangle_{{\bf T}_{N}^{d},\beta}
$ {\it be given by equality} (\ref{6}). 
{\it Then, for any real smooth differential} 2 - {\it form}
$\phi $ {\it on the torus} ${\bf T}^{d}$, $d = 3,4$, {\it of radius} $R$
{\it and for any number} $g > 0$, {\it we have}
\begin{equation}
\label{20}
\lim_{N \rightarrow \infty } \langle \exp [i (\partial \phi_{N,b} ,
\theta )] \rangle_{{\bf T}_{N}^{d} ,g^{- 2}(2\pi R)^{- d}N^{d + 2b}} =
\exp [- {g^{2} (d^{\ast }\phi ,G(d^{\ast }\phi ))}/2].
\end{equation}

It is important to note that there is no "canonical" scaling which
is used to relate the lattice theory to the continuum theory. By
taking a number $b = 1,2$,... in Propositions 1 - 3, we obtain the
different systems of natural observables and the different scalings 
which give the unique non-degenerate continuum limit. We believe that
this conclusion is valid for the lattice non-abelian gauge field theory
models also.

Let us study the correlation functions (\ref{20}) and verify the 
Wilson criterion. We substitute into relation (\ref{20}) a
square summable 2-form on the torus ${\bf T}^{d}$ given by
\begin{equation}
\label{21}
\phi = \bigl[ \prod_{i = 1,2} \Theta (L_{i} - x_{i})\bigr]
h(x_{3},...,x_{d}) dx_{1}\wedge dx_{2},
\end{equation}
where $\Theta (L - x)$ coincides with the Heaviside step function on the
interval $(0,2\pi R]$ and is periodic. Here, $h(x_{3},...,x_{d})$ is a 
smooth function on the torus ${\bf T}^{d - 2}$. Definitions (\ref{3}),
(\ref{5}) and (\ref{9}) imply that $(\partial \phi_{N,b} ,\theta )$
is a sum of Wilson loop sums for the rectangles with vertices:
$(0,0,m_{3},...,m_{d})$, $([L_{1}N(2\pi R)^{- 1}],0,m_{3},...,m_{d})$,
$([L_{1}N(2\pi R)^{- 1}],[L_{2}N(2\pi R)^{- 1}],m_{3},...,m_{d})$,

\noindent
$(0,[L_{2}N(2\pi R)^{- 1}],m_{3},...,m_{d})$, which is smeared with 
density $h_{N,b}(m_{3},...,m_{d})$ into the directions orthogonal to
the rectangles planes. Here, $[r]$ is the integer part of the real number
$r$.

The right-hand side of equality (\ref{16})  coincides with
$(2\pi R)^{- d}(d^{\ast }\phi ,G(d^{\ast }\phi ))$. It follows from
equalities (\ref{8}), (\ref{16}), and (\ref{17}) that the 
substitution of the 2-form (\ref{21}) into this expression gives
\begin{eqnarray}
\label{22}
& &(d^{\ast }\phi ,G(d^{\ast }\phi )) =
(2\pi R)^{- d} 
\sum_{{\bf l} \in {\bf Z}^{d} ; l_{1}^{2} + \cdots + l_{d}^{2} \neq 0}  
\nonumber \\
& &R^{4}(l_{1}^{2} + \cdots + l_{d}^{2})^{- 1} (l_{1}^{2} + l_{2}^{2}) 
\bigl( \prod_{k = 1,2} l_{k}^{- 2}|\exp [iR^{- 1}L_{k}l_{k}] - 1|^{2}\bigr) 
|h^{\sim }(l_{3},...,l_{d})|^{2}, 
\end{eqnarray}
where we suppose that the function $l^{- 2}|\exp [iR^{- 1}Ll] - 1|^{2}$
takes its natural value $L^{2}R^{- 2}$ at the point $l = 0$. In order
to obtain a single Wilson loop integral we need to take the density
$h(x_{3},...,x_{d}) = \delta (x_{3} - a_{3}) \cdots \delta (x_{d} - a_{d})$.
Thus, $|h^{\sim }(l_{3},...,l_{d})| = 1$ and the series (\ref{22})
seems to be divergent. From a physical point of view a thin contour
($h$ is a $\delta $-density) does not differ from a smeared one.
However, by taking the different densities $h$, we obtain the different 
asymptotics of the correlation fuctions (\ref{20}). For example, if we
take $h(x_{3},...,x_{d}) \equiv 1$, then due to definition (\ref{17})
$h^{\sim }(l_{3},...,l_{d}) =
(2\pi R)^{d - 2}\delta_{l_{3},0} \cdots \delta_{l_{d},0} $. By using
our convention on the function $l^{- 2}|\exp [iR^{- 1}Ll] - 1|^{2}$,
we have
\begin{equation}
\label{23}
(d^{\ast }\phi ,G(d^{\ast }\phi )) =
(2\pi R)^{d - 4} \sum_{{\bf l} \in {\bf Z}^{2}}  
R^{4} \prod_{k = 1,2} l_{k}^{- 2}|\exp [iR^{- 1}L_{k}l_{k}] - 1|^{2} -
(2\pi R)^{d}\biggl( \frac{L_{1}L_{2}}{(2\pi R)^{2}}\biggr)^{2}.
\end{equation}
The series in the right-hand side of equality (\ref{23}) is the
series of squared Fourier coefficient moduli for the
function $\prod_{i = 1,2} \Theta (L_{i} - x_{i})$. It follows from the
Parseval identity that this series is equal to $(2\pi R)^{2}L_{1}L_{2}$.
This implies
\begin{equation}
\label{24}
(d^{\ast }\phi ,G(d^{\ast }\phi )) =
(2\pi R)^{d}\frac{L_{1}L_{2}}{(2\pi R)^{2}}
\biggl( 1 - \frac{L_{1}L_{2}}{(2\pi R)^{2}}\biggr).
\end{equation}
Therefore, equality (\ref{24}) shows that the Wilson criterion \cite{1}
for the confinement is fuifilled for $L_{i} \leq \pi R$, $i = 1,2$
and for this particular choice of the density $h \equiv 1$. It is
evident now that the Wilson criterion does not work in the 
lattice $U(1)$ gauge theory with the periodic boundary conditions for
which it was invented. 

It is proved in the paper \cite{4} that for the differential 2-form
$\phi $ such that a restriction of any coefficient
$\phi_{ij} (x_{1} + \pi R,...,x_{d} + \pi R)$ on 
$(- \pi R,\pi R)^{\times d}$ is independent of the radius $R$ of a
torus and has a compact support, we have
\begin{eqnarray}
\label{25}
& &\lim_{R \rightarrow \infty }\lim_{N \rightarrow \infty } 
\langle \exp [i (\partial \phi_{N,b} ,
\theta )] \rangle_{{\bf T}_{N}^{d} ,g^{- 2}(2\pi R)^{- d}N^{d + 2b}} =
\nonumber \\
& &\exp \bigl[ - ({g^{2} }/2) (2\pi )^{- d} \int_{{\bf R}^{d} } d^{d} p
(p_{1}^{2} + \cdots + p_{d}^{2} )^{- 1}  \sum_{\mu = 1}^{d}
| ( d^{\ast } \phi )_{\mu }^{\sim }({\bf p}) |^{2} \bigr],
\end{eqnarray}
where the operator $d^{\ast }$ is defined by equality 
(\ref{8}) and a function $f_{\mu }^{\sim }({\bf p})$ is an usual
Fourier transform of a function $f_{\mu }({\bf x})$ on the Euclidean
space ${\bf R}^{d}$. The right-hand side of equality (\ref{25})
is a correlation function of the ${\bf R}$-gauge Euclidean 
electrodynamics \cite{6}. The correlation functions for the Euclidean
electrodynamics enable us to compute the Schwinger functions. Via the
usual analytic continuation we find \cite{6} the Wightman distributions.
These distributions define also a new Wightman theory without the 
first axiom: the test functions form a certain subspace of the space 
$S({\bf R})^{\times k}$. By an extension of these distributions to those
over all test functions from $S({\bf R})^{\times k}$ we obtain \cite{6}
the well known Gupta - Bleuler formalism for a free electromagnetic
field. Therefore, our continuum limit for lattice $U(1)$ gauge 
theory is physically reasonable.

\end{document}